# Mechanisms in Impact Fragmentation

F. K. Wittel*, H. A. Carmona**, F. Kun*** and H. J. Herrmann*

*Computational Physics for Engineering Materials IfB, Schafmattstr. 6, HIF, ETH Zurich, CH-8093 Zurich, Switzerland

** Centro de Ciências e Tecnologia, Universidade Estadual do Ceará, 60740-903 Fortaleza, Ceará, Brazil

***Department of Theoretical Physics, University of Debrecen, P.O. Box:5, H-4010 Debrecen, Hungary

+41 44 633 2871

+41 44 633 1147

fwittel@ethz.ch

http://comphys.ethz.ch

The brittle fragmentation of spheres is studied numerically by a 3D Discrete Element Model. Large scale computer simulations are performed with models that consist of agglomerates of many spherical particles, interconnected by beam-truss elements. We focus on a detailed description of the fragmentation process and study several fragmentation mechanisms involved. The evolution of meridional cracks is studied in detail. These cracks are found to initiate in the inside of the specimen with quasi-periodic angular distribution and give a broad peak in the fragment mass distribution for large fragments that can be fitted by a two-parameter Weibull distribution. The results prove to be independent of the degree of disorder in the model, but mean fragment sizes scale with velocity. Our results reproduce many experimental observations of fragment shapes, impact energy dependence or mass distribution, and significantly improve the understanding of the fragmentation process for impact fracture since we have full access to the failure conditions and evolution.

*Fragmentation; Computer Simulation; Discrete Element Method; Comminution;*

## Introduction

Fragmentation is ubiquitous in nature and can be found on all scales. Technologically we make strong use of fragmentation for example in industrial comminution processes where the focus lies on the specific reduction of material to preferred sizes, minimizing the necessary energy and amount of nano-toxic powder production. Therefore, predicting the resulting fragment mass distributions, understanding the underlying fragmentation mechanisms and scaling relations is an important field of research that has attracted the attention of researchers over the last decades. Fragmentation of single brittle spheres has been studied experimentally and numerically to understand the elementary fragmentation processes that govern comminution. Experiments from the 60s analyzed the fragment mass and size distributions (Arbiter et al. 1969; Gilvarry and Bergstrom 1961/62) with the striking observation, that the mass distribution in the range of small fragments follows a power law with exponents that are universal with respect to material or the way energy is imparted to the system. Later it was found that the exponents depend on the dimensionality of the object (Turcotte 1986). These findings were confirmed by numerical simulations, mainly based on Discrete Element Models (DEM) (Åström et al. 2000; Diehl et al. 2000; Kun and Herrmann 1996a/b). For large fragment masses, deviation from the power-law distribution could be modelled by an exponential cut-off, and by using a bi-linear or Weibull distribution (Antonyuk et al. 2006; Potapov and Campbell 1996; Cheong et al. 2004; Lu et al. 2002; Meibom and Balslev 1996; Oddershede et al. 1993). It is an every day experience that fragmentation is only obtained above a certain material dependent energy input. Numerical simulation could prove a phase transition at a critical energy with the fragmentation regime above and the fracture or damaged regime below a critical point (Behera et al. 2005; Kun and Herrmann 1999; Thornton et al. 1999). The fragmentation process itself became accessible with the availability of high speed cameras (Andrews and Kim 1998-99; Antonyuk et al. 2006; Chau et al. 2000; Majzoub and Chaudhri 2000; Salman et al. 2002; Schubert et al. 2005; Tomas et al. 1999; Wu et al. 2004). Below the critical point only slight damage could be observed, while above the specimen breaks into a small number of fragments of the shape of wedges, formed by meridional fracture planes, and additional cone-shaped fragments at the specimen-target contact point. By meridional we mean along a meridian, or other words from south to north or small to large z values. Way above the critical point, oblique fracture planes develop, that further fragment the wedge shaped fragments.

Today the mechanisms involved in the initiation and propagation of single cracks are fairly well understood, and statistical models have been successfully applied to describe macroscopic fragmentation (Åström 2006; Herrmann and Roux 1990). However, when it comes to complex fragmentation processes with instable dynamic growth of many competing cracks in the three-dimensional space (3D), much less is understood. Today model sizes become possible that allow for 3D simulations with many thousand particles and interaction forces that are more realistic than simple central potentials. These give a good refined insight of what is really happening inside the system, and how the predicted outcome of the fragmentation process depends on system properties. Numerical simulations can recover some of these findings, but while 2D models are incapable of reproducing the meridional fracture planes (Behera et al. 2005; Khanal et al. 2004; Kun and Herrmann 1999; Potapov et al. 1995/97; Thornton et al. 1996), 3D simulations were restricted to relatively small systems, and could not study the mechanisms that initiate and drive meridional fracture planes (Potapov and Campbell 1996, Thornton et al. 1999). (Arbiter et al. 1969) argued, based on high speed photographs that fracture initiates from the periphery of the contact disc between the specimen and the plane, due to the circumferential tension induced by a highly compressed cone driven into the specimen. However, their experiments did not allow access to the internal damage developed inside the specimen during impact. Using transparent acrylic resin, (Majzoub and Chaudhri 2000; Schönert 2004) observed damage initiation at the border of the contact disc, but plastic flow and material imperfections complicated the analysis. Therefore, meridional crack initiation and propagation is not fully clarified, although the resulting wedge-shaped fragments are observed for a variety of materials and impact conditions (Arbiter et al. 1969; Khanal et al. 2004; Majzoub and Chaudhri 2000; Wu et al. 2004).

In this paper we present 3D Discrete Element simulations of brittle solid spheres impacting a hard planar target. We focus our attention on the processes involved in the initiation and development of fragmentation mechanisms and how they lead to different regimes in the resulting fragment mass distributions. Our results can reproduce experimental observations on fragment shapes, scaling of impact energy dependence and mass distributions, significantly improving our understanding of the fragmentation process in impact fracture due to the time evolution of the fragmentation process and stress field involved being fully accessible.

## Model description

Discrete Element Models (DEM) were first employed by (Cundall and Strack 1979) to study rock mechanics and failure in particular. Today they are applied to quasi-static, impact or explosive loading, employing elementary particles of different shapes and density, connected by different rheological cohesive, massless elements (Bićanić 2004). Newton's equations govern the translational and rotational motion of the elements. Torques and forces can arise either from particle-particle interactions, from the cohesive elements, by interaction with boundaries like elastic or rigid walls or volumetric forces. In this work a 3D implementation of DEM is employed, that represents the solid by an agglomeration of spheres of two different sizes. The sphere centres are connected by beam-truss elements that can elongate, shear, bend and torque. The total force and moment acting on an element is composed of contact forces from sphere-sphere contacts, $\boldsymbol{F}^c = \boldsymbol{F}^{ov} + \boldsymbol{F}^{diss}$, and the stretching and bending forces $\boldsymbol{F}^b = \boldsymbol{F}^{elo} + \boldsymbol{Q}$ and moments $\boldsymbol{M}^b$ transmitted by intact beams.

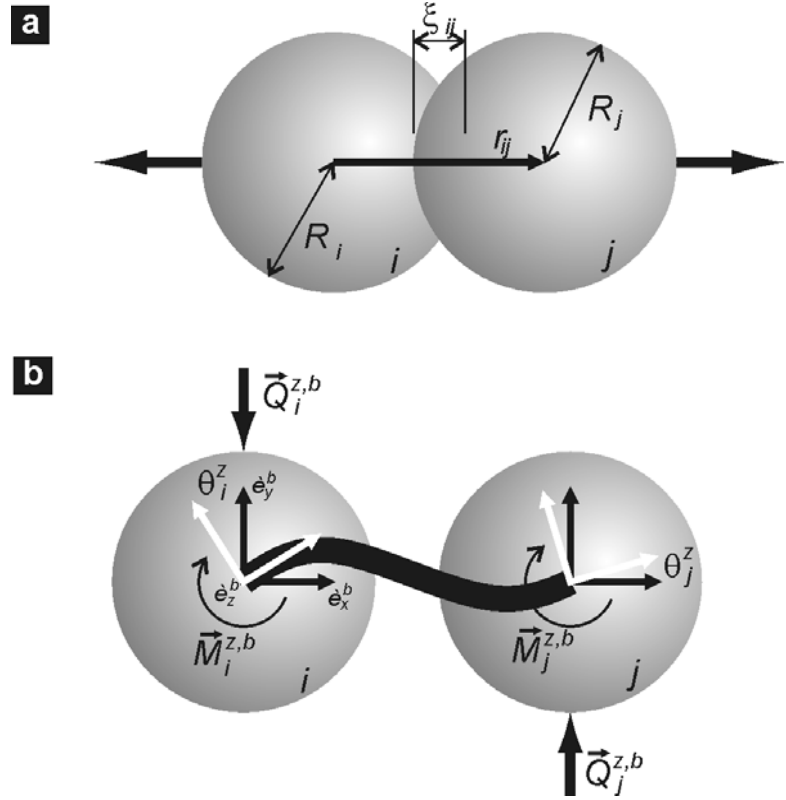

Figure 1: (a) Overlap interaction between two elements. (b) Beam deformation in the beam *x-y* plane, showing the resulting bending, shear forces and torques.

The contact force is calculated as a Hertz contact by the overlapping distance $\xi$ of spheres, by the Young's modulus of particle $E^p$, their Poisson ratio $\nu^p$ and radii $R_i$, $R_j$. In detail, the force on element *j* in the position $r_{ij}$ relative to element *i* (see Fig. 1(a)) is given by

$$\vec{F}_j^{ov} = \frac{4}{3} \frac{E^p \sqrt{R^{eff}}}{(1-\nu^2)} \xi_{ij}^{3/2} \vec{r}_{ij}, \qquad (1)$$

with the overlapping distance $\xi = R_i + R_j - |r_{ij}|$ as the sphere deformation, $1/R^{eff} = 1/R_i + 1/R_j$ and $r`_{ij} = r_{ij}/|r_{ij}|$. The additional terms of the contact force include damping and friction forces and torques that are included the same way as described in (Pöschel and Schwager 2005). The contact interaction between particles and boundaries is identical to particle-particle contact, only with $\xi = R_i - r_{ip}$, where $r_{ip}$ is the distance between the particle centre and the plane.

The 3D beam elements used are an extension of the 2D case of Euler-Bernoulli beams. In 3D the total deformation of a beam is calculated by the superposition of elongation, bending and shearing in two different planes and torsion. The force acting on element $j$ connected to element $i$ due to the elongation $\varepsilon$ of the connecting beam is given by

$$\vec{F}_j^{elo} = -E^b A^b \varepsilon \vec{r}`_{ij}, \quad (2)$$

with beam stiffness $E^b$, $\varepsilon = (|r_{ij}| - l_0)/l_0$, with the initial length of the beam $l_0$ and the beam cross section $A^b$ defined by the initial elements positions during the model construction. The flexural forces and moments transmitted by a beam are calculated from the change in the elements orientations on each beam end relative to the body-fixed $\grave{e}_x^b$- $\grave{e}_y^b$- $\grave{e}_z^b$ coordinate system of the beam. Figure 1(b) shows a typical deformation due to a rotation of both ends of the beam relative to the $\grave{e}_z^b$-axis, with $\grave{e}_x^b$ oriented in the direction of $r`_{ij}$. Given the angular orientations $\theta_i^z, \theta_j^z$, the bending force and moment $Q_j^{z,b}, M_j^{z,b}$ for the elastic deformation of the beam is given by

$$\vec{Q}_j^{z,b} = 3E^b I \frac{\left(\theta_i^z + \theta_j^z\right)}{L^2} \grave{e}_y^b, \quad \vec{M}_j^{z,b} = E^b I \frac{(\theta_i^z - \theta_j^z)}{L} \grave{e}_y^b + \left(\vec{Q}_j^{z,b} \times |\vec{r}_{ij}| \grave{e}_x^b\right), \quad (3)$$

with the moment of inertia $I$. Corresponding equations are employed for rotations around $\grave{e}_y^b$, and the forces and moments are superimposed while additional torsion moments are added for a relative rotation of the elements around $\grave{e}_x^b$,

$$\vec{M}_j^{x,b} = -G^b I^{tor} \frac{\left(\theta_j^x - \theta_i^x\right)}{L} \grave{e}_x^b, \quad (4)$$

with $G^b$ and $I^{tor}$ representing the elasticity and moment of inertia of the beams for torsion, respectively. The element forces and moments are superimposed in the global coordinate system.

To explicitly model damage, fracture and failure of the solid, beam elements are allowed to fail by a breaking rule that takes breaking due to stretching and bending of a beam into account (Herrmann et al. 1989), namely

$$\left(\frac{\varepsilon}{\varepsilon_{th}}\right)^2 + \frac{\max\left(|\theta_i|,|\theta_j|\right)}{\theta_{th}} \geq 1, \quad (5)$$

with the longitudinal strain $\varepsilon = \Delta l/l_0$ and the general rotation angles $\theta_i$ and $\theta_j$ of the beam ends and using $\varepsilon_{th}$ and $\theta_{th}$ as the respective threshold values. Note that Eq. (5) has the form of the von Mises plasticity criterion. The threshold values are taken randomly for each beam, according to the Weibull distributions

$$P(\varepsilon_{th}) = \frac{k}{\varepsilon_0}\left(\frac{\varepsilon_{th}}{\varepsilon_0}\right)^{k-1}\exp\left(-\left(\frac{\varepsilon_{th}}{\varepsilon_0}\right)^k\right), \quad P(\theta_{th}) = \frac{k}{\theta_0}\left(\frac{\theta_{th}}{\theta_0}\right)^{k-1}\exp\left(-\left(\frac{\theta_{th}}{\theta_0}\right)^k\right). \quad (6)$$

Here $k$, $\varepsilon_0$ and $\theta_0$ are model parameters, controlling the width of the distributions and the average values for $\varepsilon_{th}$ and $\theta_{th}$ respectively. Low disorder is obtained by using large $k$ values, large disorder by small $k$.

The time evolution of the system is followed solving the equations of motion for the translation and rotation of all elements using a 6$^{th}$-order Gear predictor-corrector algorithm. The dynamics of the particle rotations is described using quaternions (Rapaport 2004). The breaking rules are evaluated in each time increment. The beam breaking is irreversible, and broken beams are excluded from the model for consecutive time steps.

**Model construction and calibration**

In order to avoid artefacts arising from the system topology, like anisotropy, leading to non uniform wave propagation or preferred crack paths, special attention is given to the model construction. We first start using 27000 spherical elements that are initially placed on a large regular cubic lattice but with random velocities to randomize the system. The elements are bi-disperse in size with equal portions of $D_{min}=0.95D_{max}$. After some randomization time, a central potential, located in the centre of the simulation box, is imposed to compact the elements. The system is evolved until all particle velocities are reduced to nearly zero due to small dissipative forces. We end up with a random, nearly spherical agglomerate of particles that now get connected by beam-truss elements through a Delaunay triangulation. Note that not only contacting particles are connected. We examined the topology by looking at the angular correlations with neighbors and found no proof with respect to crystallization. After the elements are initiated, their Young's modulus is slowly increased while the centripetal gravitational field is decreased, leading to an expansion of the system. Finally the bond lengths and orientations are reset so that no initial residual stresses are present in the system and the system is trimmed to the desired shape by element removal. The beam lattice is equivalent to a material discretisation using a dual Voronoi tessellation of the domain (Bolander and Sukumar 2005; Lilliu and Van Mier 2003; Yip et al. 2006). The microscopic properties like the elastic and failure properties of elements and bonds are calibrated to obtain the desired macroscopic Young's modulus, Poisson's ratio and strength (values see Tab.1).

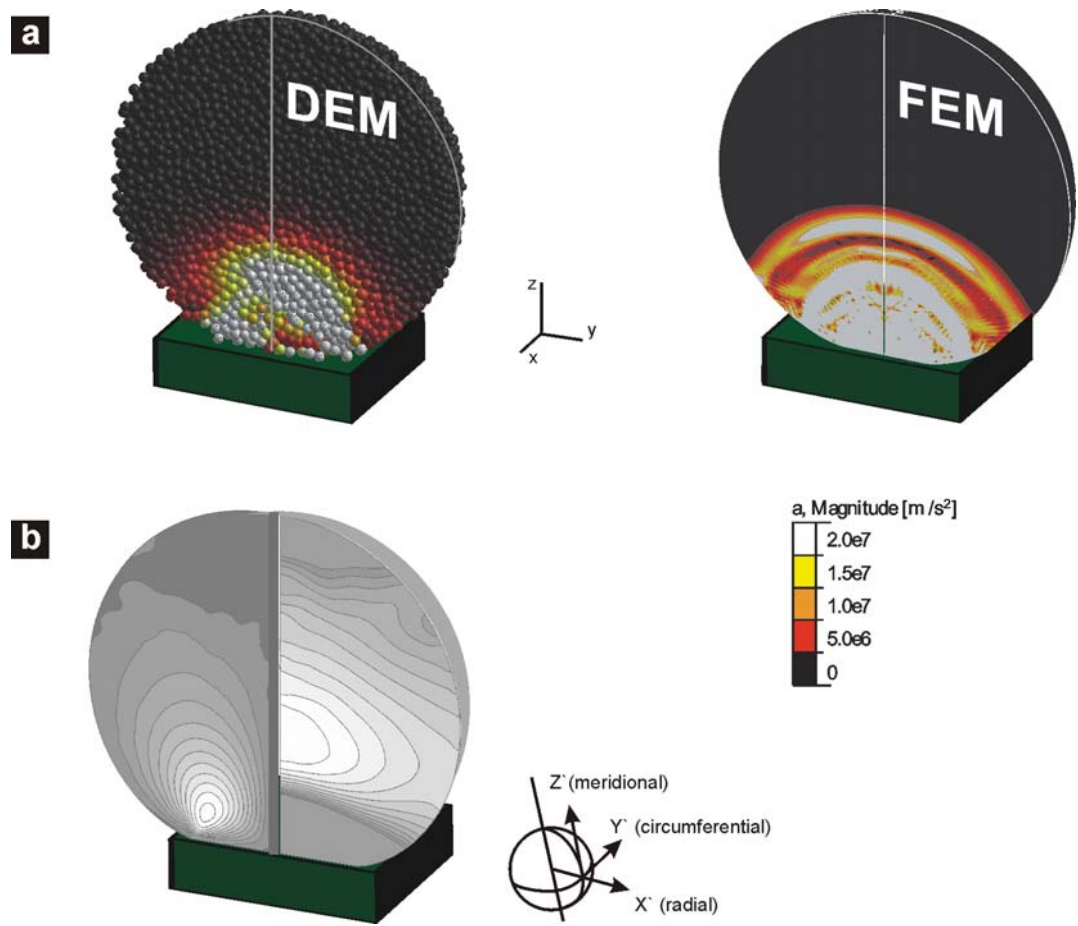

Figure 2: Cross sections of the half sphere. (a) Shock-wave propagation obtained from DEM and FEM simulations for $v_i$=117 m/s. Elements are coloured according to their acceleration magnitude. (b) Stress fields from the continuum model. The left side shows the shear stresses in global coordinates from 0 to 400 MPa (black to white) while on the right side, circumferential stresses in spherical coordinates are given, ranging from 0 to 130 MPa (black to white). Note the direction of impact corresponds to the meridional direction.

The trimmed spherical specimen is located close to a plate and an impact velocity $v_i$ in the negative $z$-direction is assigned to the system. The computation continues until no breaking activity is registered for 50 μs.

For comparative reasons we calculate the evolution of the stress field by an explicit Finite Element (FE) analysis with ABAQUS. The FE model consists of axisymmetric, linear 4-node elements, using the macroscopic properties measured on the DEM sample before (see Tab. 1). Symmetry boundary conditions are defined along the central axis of the particle and rigid ground plate. Figure 2(a) compares the shock wave of the impact using our DEM and the FEM simulation. Note that the measured wave velocity of both simulations is consistent with the analytical values (see Tab.1). The time evolution of the elastic energy in the system was found to be in excellent agreement as well.

| **Particles:** | | | | **System:** | | | |
|---|---|---|---|---|---|---|---|
| stiffness | $E^p$ | 3 | GPa | time increment | $\Delta t$ | 1e-8 | s |
| diameter | $D_1$ | 0.5 | mm | number of particles | $N^p$ | 22013 | - |
| density | $\rho$ | 3 | t/m³ | number of beams | $N^b$ | 135948 | - |
| **Beams:** | | | | solid fraction | | 0.65 | |
| stiffness | $E^b/G^b$ | 6 | GPa | sphere diameter | $D$ | 16 | mm |
| average length | $L$ | 0.5 | mm | **Macroscopic properties DEM:** | | | |
| diameter | $D$ | 0.5 | mm | system stiffness | $E$ | 7.4±0.5 | GPa |
| strain threshold | $\varepsilon_0$ | 0.02 | - | Poisson`s ratio | $\nu$ | 0.2 | - |
| bending threshold | $\theta_0$ | 3 | ° | density | $\rho$ | 1920 | kg/m³ |
| shape parameter | $\kappa$ | 0.3 | - | system strength | $\sigma_c$ | 110 | MPa |
| **Hard plate:** | | | | | | | |
| stiffness | $E^w$ | 70 | GPa | **Comparison:** | | | |
| **Interaction:** | | | | | | DEM | FEM |
| friction coefficient | $\mu$ | 1 | - | p-wave speed | | 2210 | 2270 | m/s |
| damping coefficient | $\gamma_n$ | 0.25 | s⁻¹ | | | ±100 | ±20 | |
| friction coefficient | $\gamma_t$ | 0.05 | s⁻¹ | contact time | | 31.4 | 31.4 | μs |

Table 1: Typical model properties of the DEM model e.g. calibrated on a (16x8x8) mm³ sized sample in quasi-static tensile and compressive tests.

## Mechanisms in impact fragmentation

An important step in manipulating fragmentation processes is identifying and understanding the different fragmentation mechanisms in their order of occurrence (Figs. 3(a)-(d)) and with increasing impact energy (Figs. 6(a)-(f)). The first damage mechanism observed is diffuse damage in a region approximately $D/4$ from the target plane (see Figs. 3(a)). The diffusive damage occurs due to a bi-axial stress state in the *x-y* plane (see Fig. 2(b)), superimposed by compressive stress directed *z*-wards (Andrews and Kim 1998).

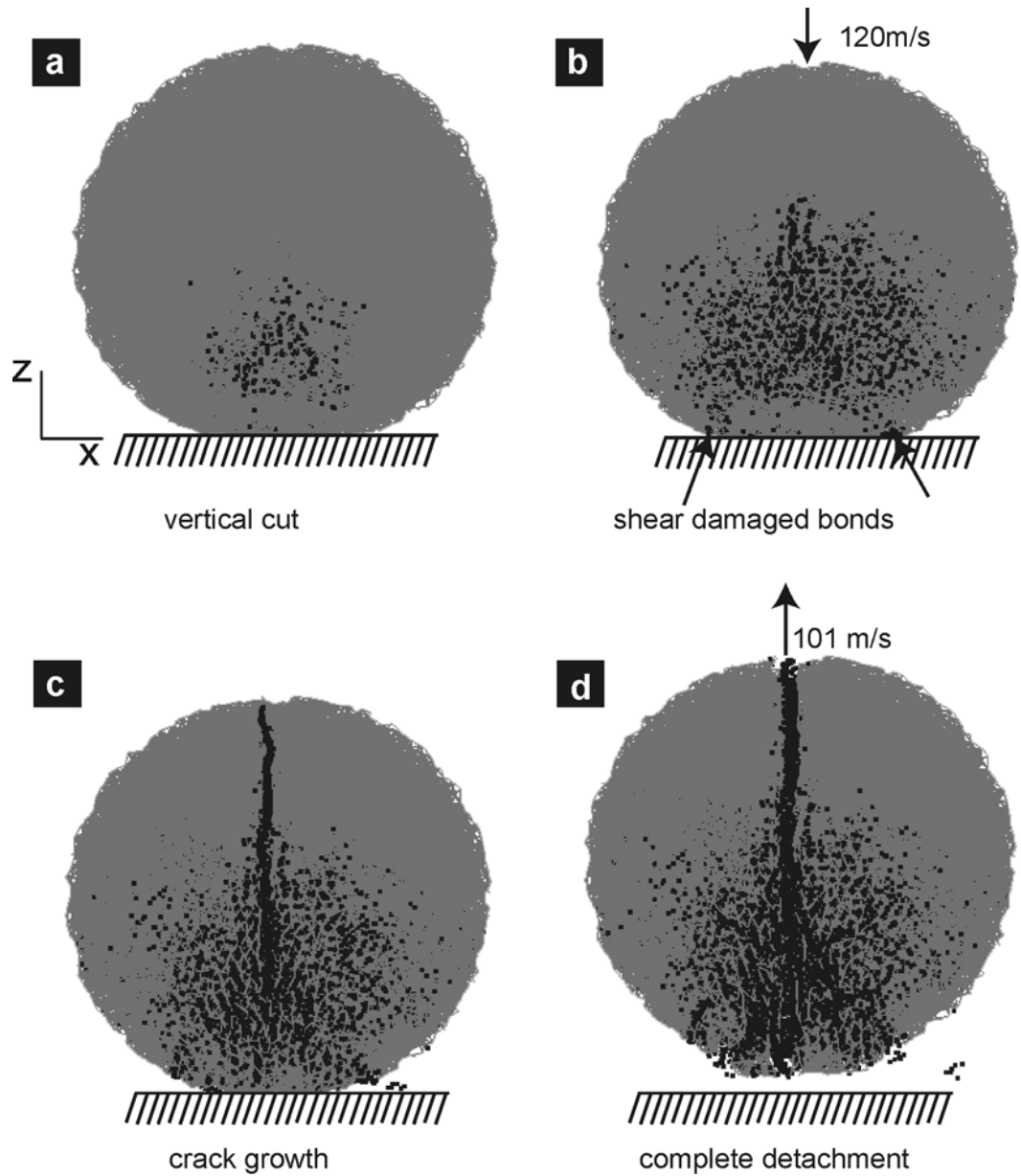

Figure 3: Simulation snapshots of vertical meridional cuts, showing broken bonds (dark colour).(a) Diffuse damage due to bi-axial stress state. (b) Formation of a ring of broken bonds by shear failure. (c) Theses broken bonds evolve into cracks that propagate inside the sample. (d) Detachment of the lower fragments.

As time evolves, meridional cracks form. Their origin is explored in Fig. 4(a), where positions and temporal evolution of the broken bonds are plotted in side and top view, showing well defined meridional crack planes that grow from the inside towards the lateral and upper free surfaces. To understand the angular separation of the crack planes, the angular distribution of the broken bonds for different times are plotted in Fig. 4(b) by using $g(\theta)$ as the probability of finding two broken bonds as a function of the angular separation $\theta$ in the *x-y* plane. The peaks in $g(\theta)$ clearly indicate meridional planes. For the velocity shown in Fig. 4, meridional cracks are separated by an average angle of about 60°, and they become evident approx. 14 µs after impact.

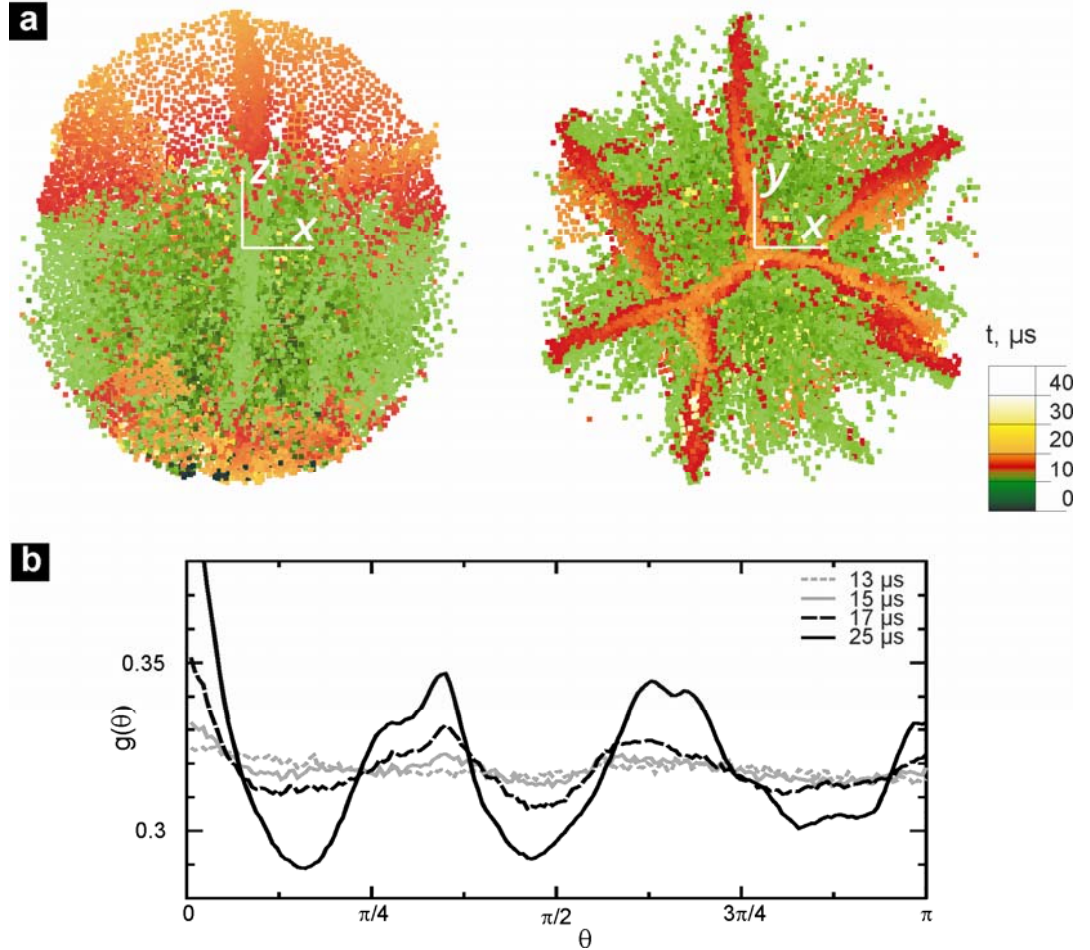

Figure 4: (a) Broken bonds coloured according to the time of failure. (b) Angular distribution functions of broken bonds ( *x-y* projection) as a function of the angular separation ($v_i$=120 m/s).

For various realizations and materials, the position of the meridional cracks changed, but not their average angular separation, even though for strong disorder (Eq. (6)), a larger amount of uncorrelated damage occurs. From the FE calculations and the damage orientation correlation plot (Fig. 4(b)) no crack orientation is preferred inside of the undamaged biaxial tensile zone. However, many micro cracks weaken this zone, decreasing its effective stiffness. Around the weakened core, the material is intact and under high circumferential tensile stress. Inside this ring shaped zone, we observe the onset of the meridional cracks when we back-trace them. With increasing impact velocity, the angular separation of crack planes decreases and thus more wedge-shaped fragments form. Obviously this effect can not be explained by arguments based on quasi-static stress analysis. However the observation can be explained in the spirit of Mott's fragmentation theory for expanding rings (Mott 1946). Once a meridional crack forms, stress is released in the neighbourhood and the stress release fronts spread with a constant velocity leading to a decreasing probability for fracture in neighbouring regions. However in the stressed regions, the strains still increase due to the external loading, and the fracture probability along with it. The average size of the wedge shaped fragments therefore is determined by the relationship between the rate at which cracks nucleate and the velocity of the stress release wave. The higher the strain rate, the higher the crack nucleation rate and the more meridional cracks are formed. Measurements of the strain rate at arbitrary positions inside the bi-axially loaded zone showed a clear correlation between impact velocity and strain rate. Even though a compact sphere and not a ring is fragmented, meridional cracks initiate in a highly stressed ring shaped region and Mott's theory can qualitatively explain the decrease of angular separation of wedge shaped fragments with increasing impact velocity. If enough energy is accessible, some of the meridional plane cracks propagate out- and upwards, fragmenting the sample into wedge shaped fragments like "orange slices".

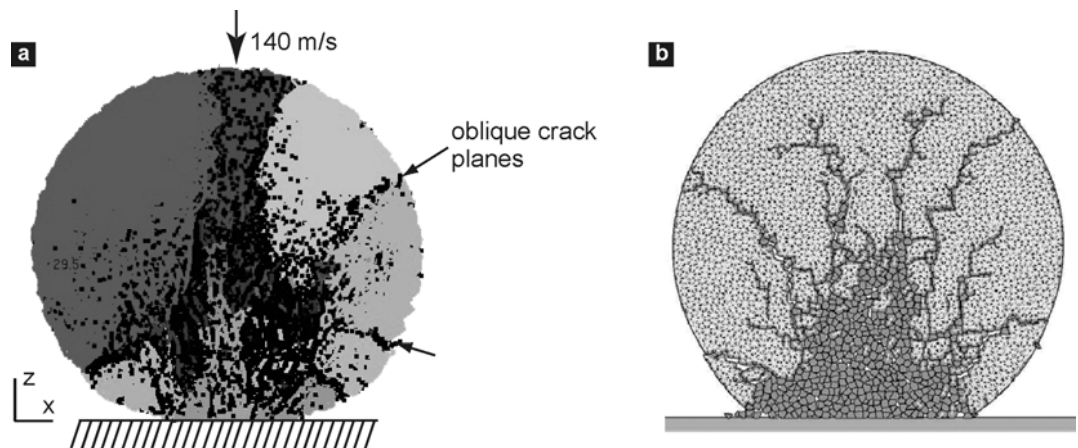

Figure 5: (a) 3D DEM simulation at $v_i$=140 m/s exemplifying the secondary cracks compared to (b) 2D simulations using polygons as elementary particles (Behera et al. 2005).

As the sphere moves further towards the plate, a ring of broken bonds forms by shear failure at the border of the contact disc (see Figs. 2(b),3). When the sample begins to detach from the plate, a cone has been formed by a ring crack that propagated from the surface to the inside of the material at approximately 45°. The resulting cone shaped fragments have a smaller rebound velocity than other fragments due to dissipated elastic energy by fracture, as can be seen in (Fig. 3(d)).

If the imparted energy is high enough, oblique plane cracks, also called *secondary* cracks may still fragment the large fragments further (see Fig. 5(a)). These secondary cracks are similar to oblique cracks observed in 2D simulations (Behera et al. 2005; Potapov et al. 1995). Fig. 5 compares the crack patterns obtained from a 2D DEM simulation that uses polygonal particles. Note that in 2D simulations, we observe an unnatural strongly fragmented cone of numerous single element fragments and meridional cracks can of course not form.

## Scaling regimes in fragmentation

For practical applications of comminution processes, the amount of energy necessary to fragment a material is an important parameter. By varying the impact energy, in fragmentation simulations and experiments two distinct regimes can be identified: below a critical energy fracture and damage takes place (Andrews and Kim 1998; Gilvarry and Bergstrom 1962; Thornton et al. 1999), while above fragments form. Figures 6(a)-(f) shows examples of final crack patterns after impact with increasing impact energy. For smaller impact energies it is possible to observe meridional cracks that reach the sample surface above the contact point, but a large piece with cracks remains (Figs. 6(a),(b)). As the initial energy increases, some of the meridional cracks are the first ones to reach the top free surface of the sphere, fragmenting the material into typically two to four fragments of wedge shape (Fig. 6(c)). Therefore meridional cracks are called *primary cracks*. As described earlier, increasing energy leads to secondary oblique plane cracks that further fragment the orange slice shaped fragments (Fig. 6(d)-(f)). The shape and number of large fragments simulated for smaller impact energies, as well as the location and orientation of oblique secondary cracks for larger energies, are in good agreement with experiments (Khanal et al. 2004; Schubert et al. 2005; Wu et al. 2004).

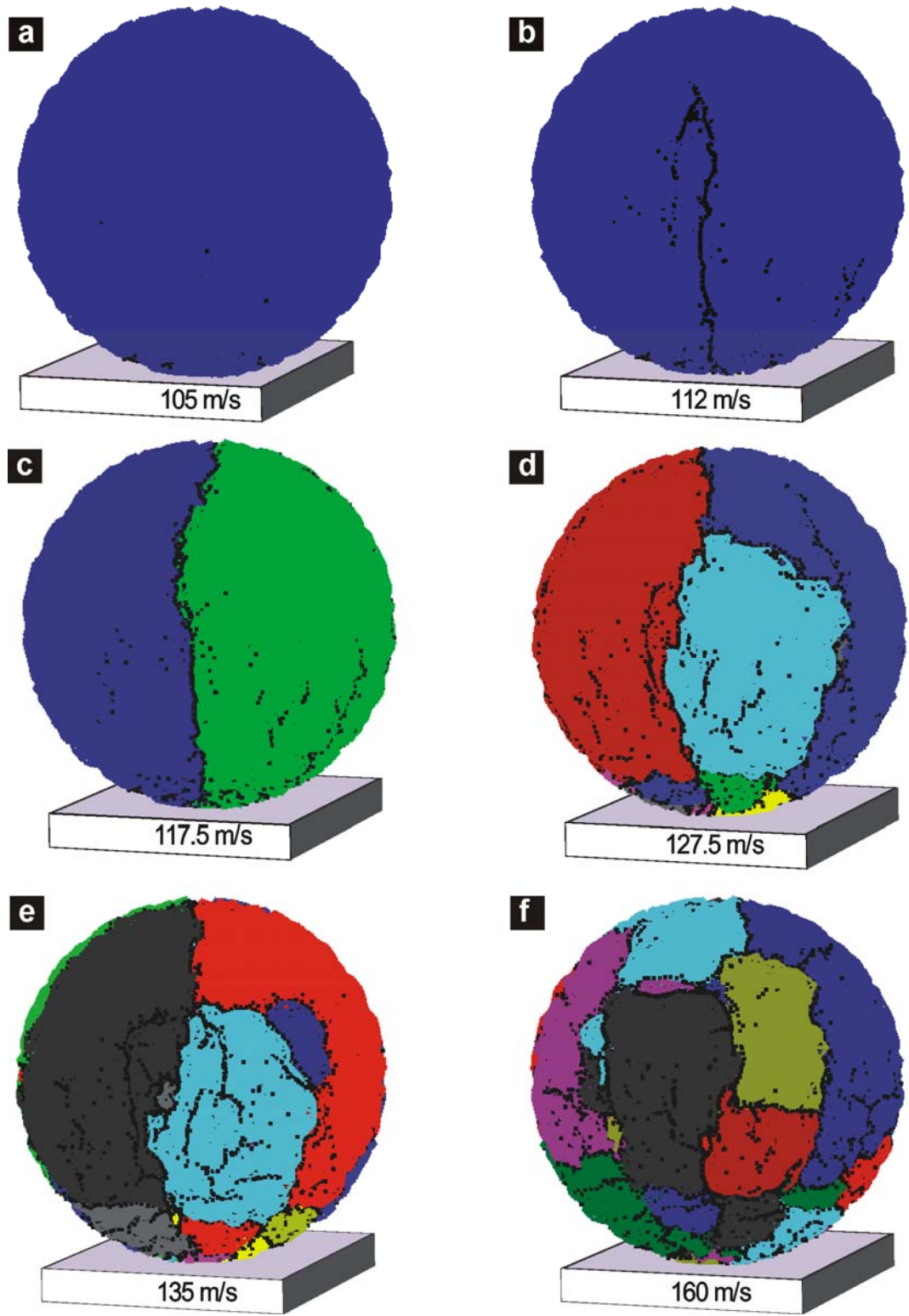

Figure 6: Final crack patterns for different initial velocities. Elements are projected to their initial positions for clearer view of the crack patterns. Intact bonds are coloured according to the final fragment they belong, while gray dots are added at the positions of broken beams.

For velocities smaller then the threshold velocity $v_{th}$, the sample is damaged by the impact but not fragmented. In particular, in 2D simulations a continuous phase transition from the damaged to the fragmented state was found (Behera et al.

2005; Kun and Herrmann 1999). Analogous to their analysis, the final state of the system is analyzed by measuring the mass of the two largest fragments, as well as the average fragment size. Note that the average mass $M_2/M_1$, with $M_k = \Sigma_i^{Nf} M_i^k - M_{max}^k$ excludes the largest fragment. We can see in Fig. 7(a) that below the threshold value $v_{th} = 115$ m/s the largest fragment in our system has almost the total mass of the system with the second largest one being consequently nearly zero. Hence, the system was only damaged and not fragmented. For $v>v_{th}$ the mass of the largest fragment rapidly decreases and the second largest and average fragment masses increase, showing a maximum at 117.5 m/s for our system. This is in very good qualitative agreement with fragmentation simulations on different geometries and loading situations (Behera et al. 2005; Kun and Herrmann 1999; Wittel et al. 2005), indicating that 3D fragmentation simulations also show this phase transition from damage to fragmented state.

**Fragment mass distributions**

Technologically the fragment mass distributions are the most important outcome of fragmentation processes. Experimental and numerical fragmentation studies show that the mass distributions follow a power law in the range of small fragments, with a universal exponent depending on the fragmentation mechanisms. The mass distribution for large fragments can be represented by an exponential cut-off of the power law. The fragment mass distribution is usually given in terms of *F(m)*, that expresses the probability density of finding a fragment with mass *m* between *m* and *m+Δm*, with *m* being the fragment mass normalized by the total system mass $M_{tot}$. The fragment mass distributions for our 3D simulations are shown in Fig. 7(b) for different impact velocities $v_i$ averaging over 36 realizations. If $v_i<v_{th}$, *F(m)* has a peak at low fragment masses corresponding to very small fragments. However the pronounced isolated peaks near the total mass of the system correspond to the large damaged, but still unfragmented system (see also Figs. 6(a),(b)). Note that fragments at intermediate mass range are not present in the damage regime. Around and above $v_{th}$, *F(m)* exhibits a power law dependence $F(m) \sim m^{-\tau}$ for intermediate masses, (dashed line in Fig. 7(b)) with $\tau = 1.9 \pm 0.2$ (Linna et al 2005; Turcotte 1986). However a local maximum can be observed for large fragments, indicating that they are formed by mechanisms that are distinct from the ones forming small fragments. The primary cracks show an angular distribution with an average separation between 45-60° resulting in fragment masses in the order of 10% of $M_{tot}$. This corresponds to the range of masses that present the broad peak in the fragment mass distribution. To give a better representation of the large fragments, the cumulative size distribution of the fragments weighted by mass, $Q_3$ is studied in Fig. 8(a). $Q_3$ is calculated by summing the mass of all fragments smaller than a given size $s$, which is estimated as the diameter of a sphere with identical mass. Note that the values are normalized by the sample diameter $D$. The shape of the size distribution for large fragments can be described by a two-parameter Weibull distribution, namely $Q_3(s) = 1 - exp[-(s/s_c)^{k_s}]$ (dashed line in Fig. 8(a), with $s_c =$ 0.75 and $k_s = 5.8$). The Weibull distribution seems suitable, since it has been empirically found to describe many fracture experiments for brittle materials (Lu et al. 2002). For increasing impact velocity $v_i$, the average fragment size decreases, which is also in agreement with experimental findings by (Antonyuk et al. 2006; Cheong et al. 2004).

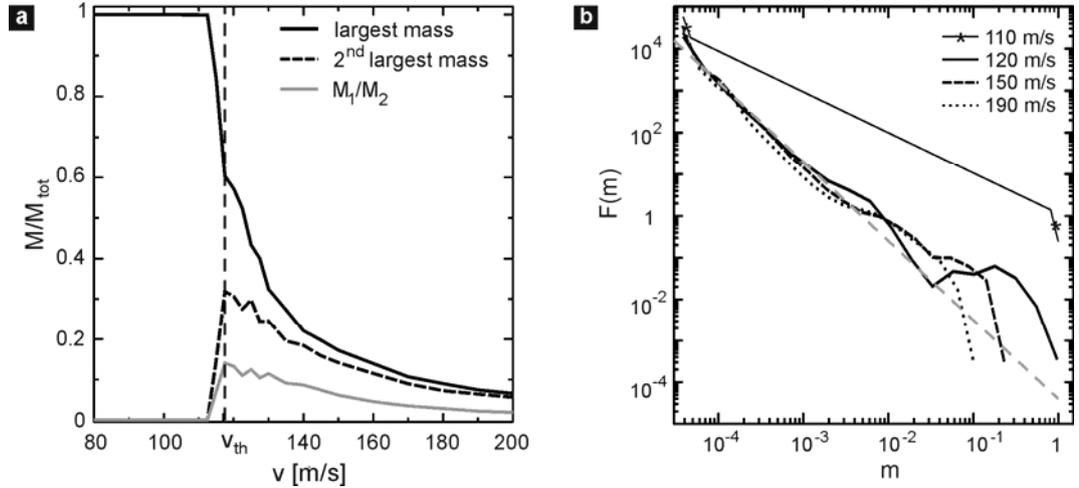

Figure 7: (a) Scaling of masses. First, second largest and average fragment mass as a function of the impact velocity. (b) Fragment mass distribution for different initial velocities. The straight line corresponds to a power-law with exponent $\tau = -1.9$.

No dependence of the fragment mass distribution with respect to material disorder ($k$ in Eq. (6)) was observed (see Fig. 8(b)). This is in agreement with observations on the universality of the fragment size distribution with respect to the breaking probability distribution (Åström et al. 2000). For the fragment mass distribution of ($v \sim v_{th}$) two distinct regimes can be identified (see Fig. 8(b)). For $m < 1/40$ (approx. 550 elements), $F(m)$ can be well described by the form

$$F(m) \sim (1-\beta)m^{-\tau}\exp\left(\frac{-m}{m_0}\right) + \beta\exp\left(\frac{-m}{m_1}\right), \qquad (7)$$

recently proposed by (Åström et al. 2004/2006). The first term is associated to branching-merging processes due to crack tip instabilities, while the second one originates from the Poissonian nucleation processes of the first dominating cracks. The parameter $\beta$ expresses the relative importance of the two processes.

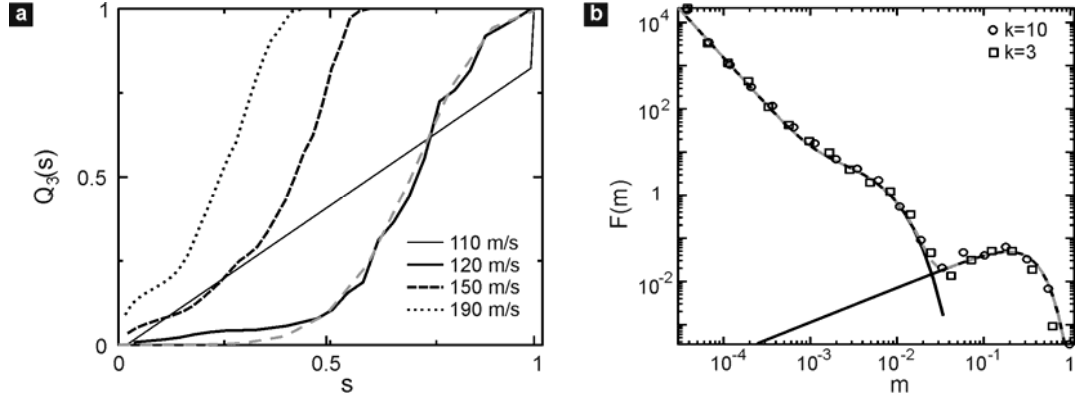

Figure 8: (a) Fragment size distribution weighted by mass for initial velocities. (b) Fragment mass distribution for $v$=122.5 m/s and different disorder in the bond breaking thresholds. The solid lines correspond to a power law with an exponential cut-off for lower masses and the Weibull distribution for large masses (Eq. (8)).

Furthermore, the scaling exponent $\tau$ only depends on the dimensionality of the system. The local maximum for the large $m$ can again be described by a two-parameter Weibull distribution

$$F(m) \sim \left(\frac{k_l}{m_l}\right)\left(\frac{m}{m_l}\right)^{k_l-1} \cdot \exp\left(-\left(\frac{m}{m_l}\right)^{k_l}\right), \qquad (8)$$

as discussed above. In Fig. 8(b), Eqs. (7),(8) are plotted separately with a dashed line, corresponding to a fit with values of $m_0$=0.001±0.001, $m_1$=0.004 ±0.001, $m_l$=0.3±0.02 and $k_l$=1.9±0.1. The good quality of the fit allows for an estimate of the exponent of the power-law distribution in the small fragment mass range to be $\tau$=2.2±0.2. This composed mass distribution function is not observed for 2D simulations (Behera et al. 2005; Kun and Herrmann 1999) or 3D simulations of shell fragmentation (Wittel et al. 2004/2005), where obviously meridional cracks are not present.

## Conclusions

We showed the importance of the use of 3D simulations for fragmentation processes by using a DEM simulation with 3D beam-truss elements for the particle cohesion. Due to this computationally more laborious approach, one is able to obtain a more realistic picture of the fragmentation processes, the evolution of fragmentation mechanisms and their consequences for the fragment mass distribution. To rationalize arguments for the fracture initiation, continuum solutions for the stress field were utilized. It was shown that 2D representations for fragmenting systems are not capable of capturing fragmentation by meridional cracks, which is the primary cracking mechanism. We showed that micro cracks form inside the sample in a region above the compressive cone long time before they are experimentally observed from the outside, if at all. They coalesce to initiate fracture in meridional fracture planes, resulting in a small number of large wedge shaped fragments. An explanation for the decrease in their angular separation could be found in the Mott fragmentation model. The resulting fragment mass distribution is described by a power law regime for small fragments and a broad peak for large fragments that can be fitted by a two-parameter Weibull distribution, in agreement with experimental results (Antonyuk et al. 2006; Cheong et al. 2004; Lu et al. 2002; Salman et al. 2002). Even though the results are valid for materials with various disorders, they are limited to the class of brittle, disordered media; however extensions to fragmentation with ductile materials are in progress. Another class of interesting questions deal with the problem of size effects, the influence of multi-disperse particles or the stiffness contrast of particles and beam-elements. For technological applications studies on the influence of target geometries and the optimization potential to obtain desired fragment size distributions or to reduce impact energies are of broad interest.

## Bibliography

Andrews EW, Kim KS (1998) Threshold conditions for dynamic fragmentation of ceramic particles. Mech. Mater. 29:161-180

Andrews EW, Kim KS (1999) Threshold conditions for dynamic fragmentation of glass particles. Mech. Mater. 31:689-703

Antonyuk S, Khanal M, Tomas J, Heinrich S, Morl L (2006) Impact breakage of spherical granules: Experimental study and DEM simulation. Chem. Eng. Process. 45:838-856

Arbiter N, Harris CC, Stamboltzis GA (1969) Single Fracture of Brittle Spheres. T. Soc. Min. Eng. 244:118-133

Åström JA, Holian BL, Timonen J (2000) Universality in fragmentation. Phys. Rev. Lett. 84:3061-3064

Åström JA, Linna RP, Timonen J, Moller PF, Oddershede L (2004) Exponential and power-law mass distributions in brittle fragmentation. Phys. Rev. E 70:026104

Åström JA (2006) Statistical models of brittle fragmentation. Adv. Phys. 55:247-278

Behera B, Kun F, McNamara S, Herrmann HJ (2005) Fragmentation of a circular disc by impact on a frictionless plate. J. Phys-Condens. Mat. 17:S2439-2456

Bićanić N (2004) Discrete Element Methods. In Stein E, deBorst R Hughes T (Edts.) Encyclopedia of Computational Mechanics: Fundamentals, Wiley and Sons, pp.311-337

Bolander JE, Sukumar N (2005) Irregular lattice model for quasistatic crack propagation. Phys. Rev. B. 71:094106

Chau KT, Wei XX, Wong RHC, Yu TX (2000) Fragmentation of brittle spheres under static and dynamic compressions: experiments and analyses. Mech. Mater. 32:543-554

Cheong YS, Reynolds GK, Salman AD, Hounslow MJ (2004) Modelling fragment size distribution using two-parameter Weibull equation. Int. J. Miner. Process. 74:S227-237

Cundall PA, Strack ODL (1979) Discrete Numerical-Model for Granular Assemblies. Geotechnique. 29:47-65

Diehl A, Carmona HA, Araripe HA, Andrade JS, Farias GA (2000) Scaling behavior in explosive fragmentation. Phys. Rev. E. 62:4742-4746

Gilvarry JJ, Bergstrom BH (1961) Fracture of Brittle Solids: 1. Distribution Function for Fragment Size in Single Fracture 32:400-410

Gilvarry JJ, Bergstrom BH (1962) Fracture of Brittle Solids: 2-Dimensional Function for Fragment Size in Single Fracture 33:3211-3213

Herrmann HJ, Roux S (eds.) (1990) Statistical Models for the Fracture of Disordered Media. North-Holland, Amsterdam

Herrmann HJ, Hansen A, Roux S (1989) Fracture of Disordered, elastic Lattices in 2 Dimensions. Phys. Rev. B. 39:637-648

Khanal M, Schubert W, Tomas J (2004) Ball impact and crack propagation - Simulations of particle compound material. Granul. Matter. 5:177-184

Kun F, Herrmann HJ (1996a) Fragmentation of colliding discs. Int. J. Mod. Phys. C 7:837-855

Kun F, Herrmann HJ (1996b) A study of fragmentation processes using a discrete element method. Comput. Method. Appl. M. 138:3-18

Kun F, Herrmann HJ (1999) Transition from damage to fragmentation in collision of solids. Phys. Rev. E 59:2623-2632

Lilliu G, Van Mier JGM (2003) 3D lattice type fracture model for concrete. Eng. Fract. Mech. 71: 927-941

Linna RP, Åström JA, Timonen J (2005) Dimensional effects in dynamic fragmentation of brittle materials. Phys. Rev. E. 72:015601

Lu CS, Danzer R, Fischer FD (2002) Fracture statistics of brittle materials: Weibull or normal distribution. Phys. Rev. E 65:067102

Majzoub R, Chaudhri MM (2000) High-speed photography of low-velocity impact cracking of solid spheres. Philos. Mag. Lett. 80:387-393

Meibom A, Balslev I (1996) Composite power laws in shock fragmentation. Phys. Rev. Lett. 76:2492-2494

Mott NF (1946) Fragmentation of Spherical Cases. Proceedings of the Royal Society of London A 189:300-308

Oddershede L, Dimon P, Bohr J (1993) Self-Organized Criticality in Fragmenting. Phys. Rev. Lett. 71:3107-3110

Potapov AV, Hopkins MA, Campbell CS (1995) A 2-Dimensional Dynamic Simulation of Solid Fracture. I. Description of the Model. Int. J. Mod. Phys. C. 6:371-398

Potapov AV, Campbell CS (1996) A three-dimensional simulation of brittle solid fracture. Int. J. Mod. Phys. C. 7:717-729

Potapov AV, Campbell CS (1997) The two mechanisms of particle impact breakage and the velocity effect. Powder Technol. 93:13-21

Pöschel T, Schwager T (2005) Computational Granular Dynamics : Models and Algorithms. Springer-Verlag Berlin Heidelberg New

Rapaport DC (2004) The Art of Molecular Dynamics Simulation. Cambridge University Press, Cambridge

Salman AD, Biggs CA, Fu J, Angyal I, Szabo M, Hounslow MJ (2002) An experimental investigation of particle fragmentation using single particle impact studies. Powder Technol. 128:36-46

Schönert K (2004) Breakage of spheres and circular discs. Powder Technol. 143-144:2-18

Schubert W, Khanal M, Tomas J (2005) Impact crushing of particle-particle compounds - experiment and simulation. Int. J. Miner. Process. 75:41-52

Thornton C, Yin KK, Adams MJ (1996) Numerical simulation of the impact fracture and fragmentation of agglomerates. J. Phys. D. Appl. Phys. 29:424-435

Thornton C, Ciomocos MT, Adams MJ (1999) Numerical simulations of agglomerate impact breakage. Powder Technol. 105:74-82

Tomas J, Schreier M, Groger T, Ehlers S (1999) Impact crushing of concrete for liberation and recycling. Powder Technol. 105:39-51

Turcotte DL (1986) Fractals and Fragmentation. J. Geophys. Res-solid 91:1921-

Wittel FW, Kun F, Herrmann HJ, Kröplin BH (2004) Fragmentation of Shells. Phys. Rev. Lett. 93:035504

Wittel FW, Kun F, Herrmann HJ, Kröplin BH (2005) Breakup of shells under explosion and impact. Phys. Rev. E. 71:016108

Wu SZ, Chau KT, Yu TX (2004) Crushing and fragmentation of brittle spheres under double impact test. Powder Technol. 143-4:41-55

Yip M, Li Z, Liao BS, Bolander JE (2006) Irregular lattice models of fracture of multiphase particulate materials. Int. J. Fracture. 140:113-124